\documentclass[preprint,showpacs,preprintnumbers,amsmath,amssymb]{revtex4}
\usepackage{epsfig}
\usepackage{graphicx}
\usepackage{amsmath}
\usepackage[english]{babel}
\usepackage{multirow}
\usepackage{placeins}

\usepackage[dvips]{color}

\begin{document}

\title{Scaling relations and finite-size scaling in gravitationally correlated lattice percolation models}


\author{Chen-Ping Zhu$^1$}
\email{oldpigman1234@126.com}
\author{Long-Tao Jia$^1$}
\author{Long-Long Sun$^1$}



\author{Beom Jun Kim$^2$}
\author{Bing-Hong Wang$^3$}
\author{Chin-Kun Hu$^{4,5}$}
\email{huck@phys.sinica.edu.tw}
\author{H. E. Stanley$^6$}

\affiliation{$^1$College of Science, Nanjing University of Aeronautics
and Astronautics, Nanjing 210016, China}

\affiliation{$^2$BK21 Physics Research Division and Department of Physics,
Sungkyunkwan University,Suwon 440-746,Korea}

\affiliation{$^3$Department of Modern Physics,University of Science and Technology of China, 200093, China}
\affiliation{$^4$Institute of Physics of Academia Sinica, Taipei 11529, Taiwan}
\affiliation{$^5$Department of Physics, National Dong Hwa University, Hualian 97401, Taiwan}

\affiliation{$^6$Center for Polymer Studies, Department of Physics, Boston University, Boston, Massachusetts 02215, USA}

\date{\today}

\begin{abstract}
In some systems, the connecting probability (and thus the percolation
process) between two sites depends on the geometric distance between them.
 To understand such process, we propose gravitationally correlated percolation models
for link-adding networks on the two-dimensional lattice $G$ with two strategies $S_{\rm max}$ and $S_{\rm min}$,
to add a link  $l_{i,j}$ to connect site $i$ and site $j$ with mass $m_i$ and $m_j$, respectively; $m_i$ and $m_j$
are sizes of the clusters which contain site $i$ and site $j$, respectively.
The probability to add the link $l_{i,j}$ is related to the generalized gravity $g_{ij} \equiv m_i m_j/r_{ij}^\emph{d}$,
where $r_{ij}$ is the geometric distance between $i$ and $j$, and $\emph{d}$ is
an adjustable decaying exponent. In the beginning of the simulation, all sites of $G$ are occupied and
there is no link. In the simulation process, two inter-cluster links $l_{i,j}$ and $l_{k,n}$ are randomly chosen
and the generalized gravities $g_{ij}$ and $g_{kn}$ are computed. In the strategy $S_{\rm max}$,
the link with larger generalized gravity is added. In the strategy $S_{\rm min}$,
the link with smaller generalized gravity is added, which include percolation on the Erd\H os-R\'enyi random graph
and the Achlioptas process of explosive percolation as the limiting cases, $d \to \infty$ and $d \to 0$, respectively.
Adjustable strategies facilitate or inhibit the
network percolation in a generic view. We calculate percolation thresholds $T_c$ and critical exponents $\beta$ by numerical simulations.
We also obtain various finite-size scaling functions for the node fractions in percolating clusters or arrival of saturation length with
different intervening strategies.
\end{abstract}


\maketitle

\section{Introduction}

The study of phase transitions and critical phenomena has attracted much attention in recent decades
\cite{71Stanley,14cjpHu,18cjpIsing,19preDimer,19cjpIsing}. The key concepts in such studies include
critical point, critical exponent, universality, scaling, and finite-size scaling function
\cite{84prbPF,94jpaHu,95prlHu,96prlHu,98preLinHu,99preOkabe}.
In this paper, we will address the problem of critical behavior of network percolation.

Network percolation has been playing an important
role as a simplified model to understand spreading processes of
message, disease, matter and dynamic processes in complex
systems~\cite{percolation1,percolation2,percolation3,percolation4,PRR2019,Newman,Bianconi,Radicchi2015,Allard}.
It has been attracting more and more
attention from physics and other research communities. With the
paradigm of complex networks, nodes representing individuals and
links interactions among them, percolation in networks serves as a
bridge connecting classical model of statistical physics and
practical problems in various fields~\cite{Achlioptas}. However,
further application of the theory is somewhat limited  since links
in networks are often in the sense of topology, i.e., connecting
relations without taking into account the geometric distance. By contrast, it is
necessary to have a geometric controllability in network
percolation, i.e., to facilitate or inhibit network
percolation in link-adding processes based on the geometric distance,
which motivates us to free ourselves from the constraint of purely
topological connection between nodes in previous models. As the consequence,
intervening strategies for this kind of correlated percolation \cite{Weinrib,PhysRept2017,Cherag,rigid}
lead to new scaling relations and finite-size scaling.

In some systems, the connecting probability (and thus the percolation
process) between two sites depends on the geometric distance between them. Mobile ad
hoc network (MANET)~\cite{MANET}, as an example, is a new wireless
communication system for temporal assembly of moving members. The
flooding mechanism \cite{MANET} of its message pervading can be viewed as a
percolation process. A MANET should assign proper transmission
range~\cite{MANET,global connectivity,epidemics2} for all nodes to prevent interference among
themselves,
and to save energy for longer lifetime of the network since they
could not be recharged during motion. Therefore, direct
communications can happen only inside speaking nodes' transmission
circles~\cite{Wangli2,epidemics2}, outside which nodes are linked in a manner of multi-hop
(indirect wireless connections through successive relays). Here,
global connectivity~\cite{global connectivity} relies on a suitable
design of transmission range adapting to the occupation density of
nodes on a two-dimensional (2D) plane. Besides, the traffic flux and
bilateral trade volumes between two cities or countries are found to
be proportional to the gross economic quantity of each side, and
inversely proportional to the distance between them. Therefore,
gravitation models ~\cite{gravitation1,gravitation2} are often used to understand
empirical data in various situations. The spread of the ground traffic
congestion could be viewed as another kind of distance-related
percolation in which Manhattan distance (the summation over
projected lengths of geometric distance along two perpendicular
directions) plays a key role. Therefore, Li, et al.~\cite{Li} pointed
out that a power-law distance-decaying link-adding probability in a
2D lattice could optimize ground traffic under certain constraints
on total cost. Moreover, a disaster gravity
mobility model~\cite{gravitation ad hoc} for MANET defines a maximum
distance at which an event affects objects in a gravitational style. That is
why pervasive disasters or rush-hour congestion can cause
percolation-like phenomena between objects \cite{pervasive and rushhour}.
In short, to properly
understand percolation in some real networks, we should not ignore
linking effect related to the geometric distance.

In practice, people often need to combine percolation process with
strategies to achieve better results of coevolutionary processes. In
the situation of massive disaster, base stations of mobile
communication often suffer from black-out, yielding a large scale of
disconnected population. In order to deliver messages, energy, and
matter supplies in disaster relief efforts to all panicked people as
soon as possible, one needs to facilitate percolation in link-adding
networks of vehicles equipped with MANET nodes or other systems.
While in other situations, such as the spread of ground traffic
congestion and epidemics \cite{epidemics1,epidemics2} which depend on the geometric
distance, one
should design effective measures to inhibit percolation. One possible
algorithm for such inhibition is the product rule (PR)
proposed by Achlioptas, et al.~\cite{Achlioptas} and other models
suggested afterwards~\cite{2,3,4,16,5,6,7,8,9,10,11,12,13,14,15,17,18,19,Cho2013,Ziff2012}.
 Original PR starts from a network with isolated nodes
 as the initial condition. During the evolution process, a
node $i$ is labeled by its mass (or called size) $m_i$ which is the
number of connected nodes in the cluster that includes node $i$.
Two topological links are randomly put into the set of the nodes at
every time step, and only the one connecting two nodes $i$ and $j$
with smaller product of masses ($m_i m_j$) is retained. This rule
postpones the development of the giant component, and a sharp change
of the fraction of the nodes in the largest cluster is observed, which has been called
 ``explosive percolation". Instead of investigating
the nature of such an unusual continuous ~\cite{Dorogov, 2011science, BJKim,Grassberg}
or discontinuous \cite{Cho2013,Ziff2012} phase transition, we are concerned with
how to facilitate or inhibit percolation in a kind of extended scheme in network growth process.

In this paper, we propose several percolation schemes on a 2D plane with
link-adding rules depending on the geometric distance, which takes the
form similar to Newton's gravity. Simply by adopting the strategy of
either maximum or minimum gravity in successive linking steps for different cases,
one can facilitate or inhibit network percolation in a systematic
way. The observed size of the largest component (cluster), and  average connection
lengths of various link types, are revealed to follow
scaling relations which were not recognized in purely
topological percolation models. The present scheme gives a generic
picture for percolation processes in real systems which are often
inevitably geometrically constrained.

\section{Models}

Suppose $N$ isolated nodes are uniformly scattered on a two-dimensional (2 D)
plane with the edge length $L$, hence $N = L^2$. For the convenience to
calculate distance, the plane is discretized with a triangular lattice $G$. Each vertex of the triangles
is occupied by a node.

As in product rule (PR) of Achlioptas process~\cite{Achlioptas} , we pick randomly two pairs [$(i,j)$ and
$(k,l)$] of nodes in the plane at every time step. For the pair
$(i,j)$ (and for $(k,l)$ likewise), we compute the generalized
gravity defined by $g_{ij} \equiv m_i m_j/r_{ij}^\emph{d}$, where
$m_i$ and $m_j$ are the number of sites of the clusters which include site $i$ and site $j$, respectively,
$r_{ij}$ is the geometric distance between $i$ and $j$, and $\emph{d}$ is
an adjustable decaying exponent. Once we have $g_{ij}$ and
$g_{kl}$, we have two choices in selecting which link should be retained.
For the case of the maximum gravity strategy (we call it $S_{\rm
max}$) we connect the pair with the larger value of the gravity,
e.g., the link $(i,j)$ is made if $g_{ij} > g_{kl}$, and the link
$(k,l)$ otherwise. We also use the minimum gravity strategy
($S_{\rm min}$) in which we favor the pair of nodes with smaller gravity to make
connection. The two strategies, $S_{\rm max}$ and $S_{\rm min}$,
lead the link-adding networks to evolve along the opposite
percolation processes. Generally speaking, $S_{\rm max}$
facilitates the percolation process, whereas $S_{\rm min}$
inhibits it similar to explosive percolation
\cite{Achlioptas,Cho2013,Dorogov, 2011science,BJKim}. All such generalized gravity values are
calculated inside the circular transmission range with the radius
$R$ centered at one of nodes $i$ and $j$ as the speaking
node~\cite{Wangli2,epidemics2} in a MANET, for example.

For the different limits of parameters $R$ and $\emph{d}$, we have
two cases. Case 1: With the transmission range
$R\rightarrow \infty$, we have a generalized gravitation rule
which is an extension~\cite{gravitation ad hoc} of widely used
gravitation model $(\emph{d} = 1)$~\cite{gravitation1,gravitation2}
with the decaying exponent $\emph{d}$ tunable.
Case 2: With both adjustable values of radius $R$ and
exponent $\emph{d}$, we have the gravity rule \cite{gravitation1,gravitation2} inside the transmission
range. It can describe the communication or traffics with
constrained power or resources.

\section{Simulation results}

 All simulations are carried out on the $L\times L$ triangular lattice of the size
 $N = L\times L$ with $L = 32, 64, 128$ and $256$, respectively. We simulate either of strategy $S_{\rm max}$ or
$S_{\rm min}$. The total number of links
equating to that of time-steps is divided by $N$, which is defined
as $T$. The mass of the largest component divided by $N$ makes up
the observable $C_{1}$, the node fraction of the largest component.
The algorithm in the present model is similar to that of
Ref.\ \cite{3} including the rule of intra-cluster priority, except
distance-decaying exponent $\emph{d}$ and transmission radius $R$ used.
And similar time-dependent variation of fractions of different
types (I: both inter-clusters; II: one inter-cluster and the other
intra-cluster; and III: both intra-clusters, see Fig. 2 in \cite{3}) of links
retained \cite{3, 19} are also observed near the threshold of
percolation. All results presented in this work are obtained from
 the average over 100 different realizations of network
configurations.



For strategy $S_{\rm max}$ in Case 1,  the percolation threshold decreases
from the limit $T_c = 0.5 $ for the  Erd\H os-R\'enyi (ER) random graph
[see Fig.~1(a)]. As the exponent $\emph{d}$ decreases, $T_c$ shifts
downward (e.g.,$T_c = 0.37$ for $\emph{d} = 0.2$ and $T_c = 0.36$ for $\emph{d} = 0.01$).
Following the standard manipulation ~\cite{Christensen}, we
obtain a group of decaying - exponent
$\emph{d}$ - dependent percolation thresholds $T_c$, and corresponding critical exponents
$\beta_1$ in the probability for a node to be in the percolating cluster:

\begin{equation}
 C_{1} \sim (T - T_c)^{\beta_1} \ \ {\rm for} \ \ T  \to  T_c +.
\end{equation}
Numerical results for $S_{\rm max}$ in Case 1 are listed in Table I. Obviously, $T_c$ and $\beta_1$ increase with $d$.
Figure 1(b) shows that for $S_{\rm min}$, $T_c$ and the exponent $\beta_1$ also depends on $d$.

In addition, another special point $T_0$ attracts our attention.
Curves $C_{1}(T)$ of Fig. 1(a) for different $\emph{d}$ cross approximately at a point
$T_0~( >T_c )$. Let $t = (T - T_0)/T_0$, then, $C_{1}(T)$ can be roughly re-scaled
as
\begin{equation}
 C_{1} \sim \emph{d}^{-\omega} f( t \emph{d}^{\epsilon})
\end{equation}
for different exponents $0.2 < \emph{d} \le 2$
(inset of Fig.~1(a), except the situation $\emph{d} = 0.2$ with a dashed green line),
 where $T_0 = 0.78$, , $\omega = 0.01$, $\epsilon = 0.20$,
and $f(x)$ is a universal scaling function, which is
similar to the super-scaling behavior studied by Watanabe and Hu~\cite{08preWH}.


With the strategy $S_{\rm min}$ in Case 1, PR ~\cite{Achlioptas} can be resumed by
letting $\emph{d} \rightarrow 0$, with the threshold $T_c$ approaching
$0.888$ which is the transition point of Achlioptas-type percolation~\cite{Achlioptas}.
On the other hand, $\emph{d} ~\rightarrow \infty$, the gravity values for both candidate links
become indistinguishable and thus any one of the two is selected
arbitrarily, which resumes the case of percolation in growing ER
random graph. Fig.~1 (b) illustrates these two limiting cases and
intermediary ones between them with $L = 128$.

According to the priority rules distinguishing candidate links into three types as shown in Figure 2 of \cite{3},
we calculated the average lengths $\emph{l}_{I},\emph{l}_{II}$ and $\emph{l}_{III}$ of
type-I, type-II and type-III links, respectively, as the summations of specific link-lengths over
corresponding numbers of such types of links.
The finally saturated average lengths of both type-II and type-III links are $\emph{l}_{0} = 131.9$ for $L = 128$ (see Fig.2).
Such saturated value is reached for $T \ge T_s=1.0$.

To find the average length $\emph{l}$ till time step $T$ with
strategy $S_{\rm max}$ in case 1, we do ensemble average on geometric lengths of
retained links under different exponents ($\emph{d} = 0.2, 0.5, 1.0, 2.0, 3.0$ and $5.0$ )
for the lattice with the edge length $L = 128$. Simulation results for three types of links \cite{3, 19} are shown in
Fig. 2. Temporal variations of normalized average lengths of type-III links \cite{3, 19} are re-scaled to collapse very well
into a single curve as shown in Fig. 3. Therefore, we get the following scaling behavior:
\begin{center}
\begin{equation}
 \emph{l}/\emph{l}_{0} \sim \emph{d}^{-\lambda} F(\emph{d}^{\tau}T)
\end{equation}
\end{center}
where $\emph{l}_{0} = 131.90$ (see Fig. 2), $\lambda =-0.001$, $\tau = 0.005$ and $F(x)$ is
a universal scaling function.



As seen in Fig.2, averaged lengthes $\emph{l}$ of both type-II and type-III links grow monotonically
until they get saturated. Actually,
they approach the saturated average length $\emph{l}_{0} =131.9$ (see Fig. 2) in $\emph{d}$-dependent paces.
 Average length $\emph{l}$ in any growth step (T)
for a smaller $\emph{d}$ is longer than those with larger $\emph{d}$, because strategy $S_{\rm max}$ favors the former links,
 and the links with a larger $\emph{d}$ starts
to be realized later on average than those with smaller $\emph{d}$ due to the same reason. Interestingly, $\emph{d}$-dependent average
lengths for each type of links have their own universal scaling functions, which are illustrated in Fig. 4 and Fig. 5,
respectively. The scaling behavior for type-II links in case 1 to arrive at saturated average length $\emph{l}_{0}$ reads:
\begin{center}
\begin{equation}
 p_{2} \sim g((T - 1.0)^{\alpha_{2}} \emph{d}^{\gamma_{2}})
\end{equation}
\end{center}
where $\alpha_{2} = - 0.35$ and $\gamma_{2} = -0.03$, respectively and $g(x)$ is a universal scaling
function valid for $0.2 \le \emph{d} \le 5.0$.
 Meanwhile, the scaling behavior for type-III links in case 1 to arrive at saturated average
 length $\emph{l}_{0}$ reads:
\begin{center}
\begin{equation}
 p_{3} \sim  S((T - 1.0)^{\alpha_{3}}\emph{d}^{\gamma_3})
\end{equation}
\end{center}
where $\alpha_{3} = - 1.0$ and $\gamma_{3} = -0.08$, respectively and $S(x)$ is a universal scaling function valid for $0.2 \le \emph{d} \le 5.0$.

In addition, the difference of average lengths between type-II and type-III
links $(\emph{l}_{II} - \emph{l}_{III})$ is exactly coherent with the difference of average fractions
between these two types of links $(F_{II} - F_{III})$ at the same $T$, which is
shown in Fig. 6. Therefore, a universal function exists for
$(\emph{l}_{II} - \emph{l}_{III})$  vs. $(F_{II} - F_{III})$ in the simulated range of \emph{d} $(0.2 \le \emph{d} \le 5.0)$.
While pure Achlioptas process~\cite{Achlioptas} does not share the same property (shown in blue line).
Obviously, Fig.3 - Fig.6 and corresponding scaling relations (formulas (3), (4), and (5)) can not be accounted as
trivial ones, since they only happen to the present schemes based on the classification in ref. \cite{3,19}.


Simulations for Case 2 reveal combined effect of transmission
range and gravitation. Following the standard manipulation in ~Ref. \cite{Christensen},
 we obtain a group of decaying - exponent
$\emph{d}$ - dependent percolation thresholds $T_c$, and corresponding critical exponents
$\beta_2$ in the probability for a node to be in the percolating cluster, with the same
form as formula (1) but different exponents $\beta_2$.

\begin{equation}
 C_{2} \sim (T - T_c)^{\beta_2} for T  ~\rightarrow T_c +
\end{equation}
Numerical results for $S_{\rm min}$ in Case 2 are listed in Table II which shows that $T_c$ and $\beta_2$ depend on $d$.

In addition, another special point $T_0$ attracts our attention.
Rough scaling relations with strategy $S_{\rm min}$ are obtained for a range of $R$ $(3 < R \le 8)$ and distance-decaying exponent
 $\emph{d}$ $(0.2 < \emph{d} < 2.0)$:

\begin{center}
\begin{equation}
 C_{2} \sim (\emph{d}/\emph{d}_0)^{-\theta} h[t(\emph{d}/\emph{d}_0)^{\phi}]
\end{equation}
\end{center}
for different \emph{d}, where $T_{0} = 1.0$, $\theta = 0.005$,
$\phi = -0.50$, $\emph{d}_0 = 0.5$, and $h(x)$ is an approximate universal scaling function. For
this scaling relation, the validation range of decaying exponent $\emph{d}$
 and transmission range $R$ need to adapt to each other, since the effect of
 a weak decay with a small exponent $\emph{d}$ would be diminished by a small enough $R$ (e.g., we must
 have $\emph{d} > 0.5$ for $R = 4$), and strong enough decay (large $\emph{d}$) would
 ruin the effect of a large $R$ (e.g., we must have $\emph{d} < 5.0$ for
 $R = 8$).
 A modest example for $R = 5$ is shown in Fig.~7(a)
 (The scaling is roughly valid for $0.2 < \emph{d} < 2.0 $ ).

  Besides, a rough scaling behavior with strategy $S_{\rm max}$ for different $R$ and
$\emph{d}$ reads:
\begin{center}
\begin{equation}
 C_{2} \sim R^{-\delta} H(t\rho^{\eta})
\end{equation}
\end{center}
for $R > 3$, where $\rho = (R - R_0)/R_0$, $R_{0} = 2$, $\eta = -0.10$,
$\delta = -0.005$, $T_0 = 1.0$, and $H(x)$ is an approximate universal scaling function, which
is shown in Fig. 7(b) (scaling is only valid for a small range $(4 < R \le 8 )$. Here, $T_0$ is
another special point where
average lengths of type-II and type-III links arrive at the same
level, and fractions of type I and III links \cite{3, 19} get a
balance, meanwhile the fraction of type II links arrives at its
summit \cite{to be published}. Also, suitable match between parameters  $\emph{d}$ and $R$
is required. Otherwise, this scaling behavior is invalid, just as the case $R = 4$
in the inset of Fig. 7(b).


Inhibitory strategy $S_{\rm min}$ in Case 2 produces the largest threshold
on a 2D plane to the best of our knowledge. Through finite-size
transformation we check the critical point $T_c$. The scaling
behaviors of node fraction $C_{2}$ and susceptibility $\chi$ defined
as $\chi \equiv [{\langle C_{2}^2 \rangle - \langle
C_{2}\rangle^2}]/N$~~\cite{4,84prbHu} are
\begin{center}
\begin{equation}
 C_2 \sim N^{-\beta / \nu} Q\left(\left(T - T_{c}\right) N^{1/\nu}\right),
\end{equation}
\end{center}
\begin{center}
\begin{equation}
 \chi \sim N^{\gamma / \nu} Z\left(\left(T - T_{c}\right) N^{1/\nu}\right),
\end{equation}
\end{center}
where $1/\nu$=0.2, $\beta$/$\nu$ =0.005, $\gamma$/$\nu$ =0.995,
and $Q(x)$ and $Z(x)$ are universal scaling functions. Therefore, a scaling law
of continuous phase transition
\begin{center}
\begin{equation}
\beta/\nu +  \gamma/\nu = 1.
\end{equation}
\end{center}
remains valid for two scaling relations for different parameter sets $(R, \emph{d})$,
which is verified well although scaling relations (7) is limited within a small range for
$S_{\rm min}$ in Case 2. Similar scaling law has been obtained
by Radicchi {\it et al.}~\cite{4} for scale-free networks but with
different sets of exponents. Therefore, the present one in Fig. 8 should be concluded into a different universality class.
 Numerical evidence of $S_{\rm min}$ in Case 2 for $R = 2$, $\emph{d} = 2.0$ with $L = 32, 64, 128$ and $256$ are shown as an example
in Fig. 8(a) and Fig. 8(b), with the percolation threshold as large as
$T_c\simeq 1.5$, as an example of $S_{\rm min}$ in Case 2. Insets of them illustrate the re-scaled results of $C_{2}$
and $\chi$ (see formulas (9) and (10)), respectively.

\section{Discussion and Conclusions}

 It should be noted that scaling relations illustrated in Fig. 1, Fig. 3, Fig. 4, Fig. 5 and Fig. 7
 (formulas (2), (3), (4), (5), (7) and (8)) are not
referring to critical points $T_c$ of pertinent percolation in specific gravitational distance
- decaying schemes. Instead, they are referring to kind of special points $T_0$ which are $\emph{d}$
governing or $(\emph{d}, R)$
 coordinately controlled, and worthy of further investigation. Among
them formula (1) around $T_0$ for $S_{\rm max}$ in case 1 is approximately valid. And formula
(5) and (6)
are valid only for properly matched sets of $R$ and $\emph{d}$. By contrast, Fig. 8 and
corresponding scaling relations,  $\emph{i.e.}$
formulas (7), (8) and (9) referring to order parameter $C_{2}$ and $\chi$ around $T_c$ are
quite solid.

The gravitationally correlated lattice percolation models (GCLPMs)  introduced in this paper 
 are new models of long-range correlated percolation
\cite{Weinrib,PhysRept2017,Cherag,rigid}, and they are in different
universality class from the existing correlation percolation model, e.g. the scaling law mentioned in \cite{Weinrib}
is violated and the PR is merged into the
bond-occupation schemes. From this viewpoint we can
understand a different saturation effect of $S_{\rm max}$ for decaying
exponent $d\geq 3.0$ in Fig. 1(b) and limited validation ranges of $\emph{d}$
for all scaling relations relevant to correlations in Case 2 with strategy $S_{\rm min}$.

Intervening schemes in the present gravitational correlated percolation have predicted rich
scaling relations.
With the link-adding network schemes depending on gravitational distance-decaying strategies
$S_{\rm max}$ or $S_{\rm min}$,
we designed different ways to facilitate or inhibit network
percolation on the 2D plane from a generic view of continuous phase
transition. The adjustable
transition threshold covers the range from $0.36$ to $1.5$ with the
present simulations, which provides an approach to tuning critical point $T_c$ precisely according
to requirement of different systems.
Moreover, the approaches to re-scale time (the number
of edges $T$) of a growing network with distance information would
reveal more critical spatiotemporal properties of co-evolutionary
processes. They could get broader applications than previous network percolation
models constrained in topological sense when parameters $\emph{d}$ and $R$
are properly selected for practical problems.

The GCLPMs introduced in this paper can inspire many interesting
problems for further studies. In the present paper, we only simulate the GCLPM on the plane triangular (pt) lattice and obtain
the finite-size scaling function only for the pt lattice.
It has been found that bond and site percolation models on the square (sq), plane triangular (pt), and honeycomb (hc)
lattices can have universal finite-size scaling functions when the aspect ratios of the sq, pt, and hc lattices
are chosen to have the relative sizes  1: $\sqrt{3}/2$: $\sqrt{3}$ \cite{95prlHu,95PhysicaAHu,96prlHu}. An argument about
why to choose such aspect ratios can be found in the Appendix C of \cite{14cjpHu}. We can simulate the GCLPM
on the sq, pt and hc lattices whose aspect ratios have the relative sizes  1: $\sqrt{3}/2$: $\sqrt{3}$ to obtain
the universal finite-size scaling functions of the GCLPM on the sq, pt and hc lattices.



The Ising model and the Potts model are important lattice models \cite{71Stanley,14cjpHu,98preIzma,82rmpPotts-WuFY,96prlChenCN,96prlPotts2}.
It has been found that the Ising model on the sq, pt and hc lattices whose aspect ratios have the relative sizes  1: $\sqrt{3}/2$: $\sqrt{3}$
can have universal finite-size scaling functions \cite{99preOkabe,97preWangHu,03preWuMC}.
It has been shown that the Ising model and the Potts model are corresponding to the 2-state and the $q$-state bond-correlated
percolation models (qBCPM) \cite{14cjpHu,84prbHu,84prbHuPotts}, respectively. The 2-state bond correlated percolation
model (2BCPM) is a special case of the qBCPM when $q=2$. The random bond percolation model
 is a special case of the qBCPM when $q=1$ \cite{82rmpPotts-WuFY}.
To simulate the qBCPM, Swendsen-Wang has proposed a Swendsen-Wang algorithm \cite{87prlSW}, which can overcome the critical slowing
down. Hu and Mak had used this algorithm to simulate the qBCPM on the sq and the simple cubic lattices \cite{89prbQPM}.
Chen, Hu and Mak had developed a FORTRAN program to simulate the qBCPM on D-dimensional hypercubic lattices \cite{91cpcQPM} based on
the Swendsen-Wang algorithm \cite{87prlSW}. In this paper, we modify the bond random percolation model to introduce the GCLPM.
In the future, we can modify the qBCPM to include the concepts from the GCLPM.
Such a model can be denoted as qBCPM-GCLPM. We can simulate the qBCPM-GCLPM 
on the sq, pt and hc lattices whose aspect ratios have the relative sizes  1: $\sqrt{3}/2$: $\sqrt{3}$ to find the
universal finite-size scalings for the qBCPM-GCLPM. We can also study whether and how the qBCPM-GCLPM can show
a first-order phase transition as parameters of the model, e.g. $q$ and $d$, are changed. 

In summary, the GCLPM introduced in this paper can inspire many interesting problems for further studies.


\begin{acknowledgments}
 C.P.Z. thanks H. Park, P. Holm, X.-S. Chen, M.-X. Liu and Z.-M. Gu for useful
discussion. C. K. H. is indebted to R. M. Ziff for a critical reading of the manuscript.
C.P.Z., L.T.J. and L.L.S. acknowledge financial support from National Natural
Science Foundation of China (NNSFC) under Grants No. 11175086,
10775071 and 11775111. B.J.K. acknowledges the support from the National
Research Foundation of Korea (NRF) grant funded by the Korea Government (MSIT)
Grant No. 2017R1A2B2005957. C.K.H. is supported by Grant MOST 108-2112-M-259 -008.
\end{acknowledgments}

{}

\newpage
\begin{table}
\caption{Critical points $T_c$ and critical exponents $\beta_1$ of $\emph{d}-$ dependent percolation probability $C_1$ for a site to be in the percolating cluster with strategy $S_{\rm max}$ and Case 1}
\label{tab:datasets}
\begin{tabular}{lllllllllllll}
\hline
\hline
        $\emph{d}~~$ &  0.01  & 0.1  &   0.2 &  0.5  &  0.8 &  1.0   & 1.2   & 1.5   &   2.0 \\

\hline
      $T_{c}~~$   & 0.356(5) & 0.358(3)  &  0.369(3) & 0.408(3) & 0.442(7) & 0.460(6)  & 0.474(9) & 0.491(0) & 0.509(0) \\
\hline
      $\beta_1 ~~$ & 0.894(8) & 0.943(7) & 0.971(1)& 0.984(9) & 1.01(1) & 0.990(3)& 1.00(6) & 1.02(7) & 1.02(6) \\
\hline
\hline

\end{tabular}
\end{table}

\begin{table}
\caption{Critical points $T_c$ and critical exponents $\beta_2$ \emph{d}- dependent percolation probability $C_2$ for
a site to be in the percolating cluster with strategy $S_{\rm min}$ and Case 2, $R = 5$.}
\begin{tabular}{llllllll}
\hline
        $\emph{d}$ &   0.2 &  0.5  &  0.8 &  1.0   &   1.2  &  1.5  &  2.0  \\
\hline
\hline
      $T_{c}$  &  0.874(8) & 0.863(7) & 0.846(6) & 0.831(7) & 0.816(8) & 0.792(0) & 0.742(6) \\
\hline
$\beta_2 $ & 1.18(2) & 1.34(4) & 1.28(0) & 1.31(9) & 1.26(0) & 1.25(0)   & 1.26(1) \\
\hline
\hline

\end{tabular}
\end{table}

\newpage

\begin{center}
\begin{figure}
\begin{center}
\includegraphics[width=0.80\textwidth]{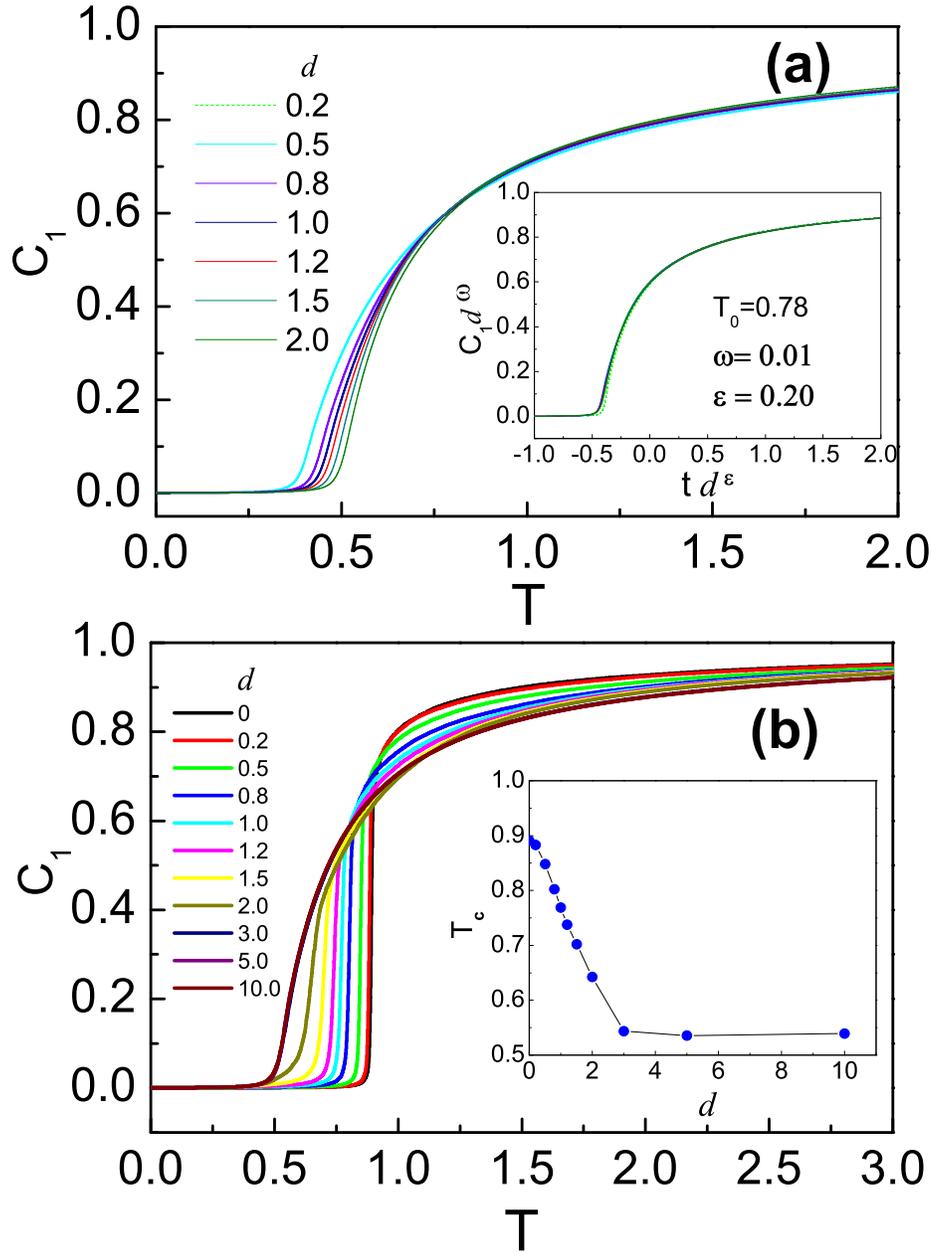}
\end{center}
\caption{(color online) Node fraction $C_{1}(T)$ of
the largest component in Case 1 ($R\rightarrow \infty$). (a) For
strategy $S_{\rm max}$. Inset: Re-scaled $C_{1}d^{\omega}$ as a function of $td^{\epsilon}$ with $T_{0} = 0.78$ and $t=(T-T_0)/T_0$.
(b) $C_{1}(T)$ for strategy $S_{\rm min}$. Inset: $T_{c}$ vs. \emph{d}. For both (a) and (b) cases, $L = 128$.}
\end{figure}
\end{center}

\begin{figure}
\begin{center}
\includegraphics[width=0.80\textwidth]{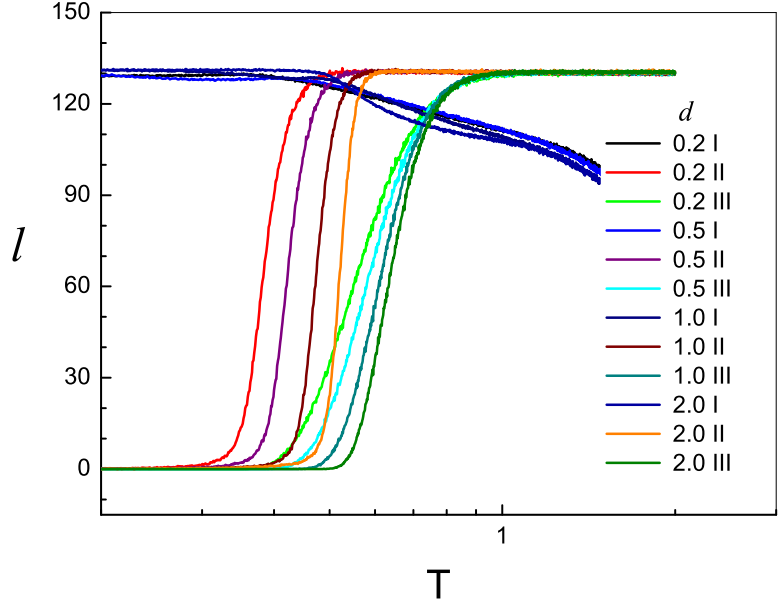}
\end{center}
\caption{(color online) The average lengths $\emph{l}$ with strategy
$S_{\rm max}$ versus time steps $T$ in Case 1 ($R \rightarrow \infty$) for different exponents
$\emph{d}$ = 0.2, 0.5, 1.0, and 2.0. In all cases, $L = 128$. The finally saturated average lengths of both type-II and type-III links
are $\emph{l}_{0} = 131.9$.}
\end{figure}

\begin{figure}
\begin{center}
\includegraphics[width=0.80\textwidth]{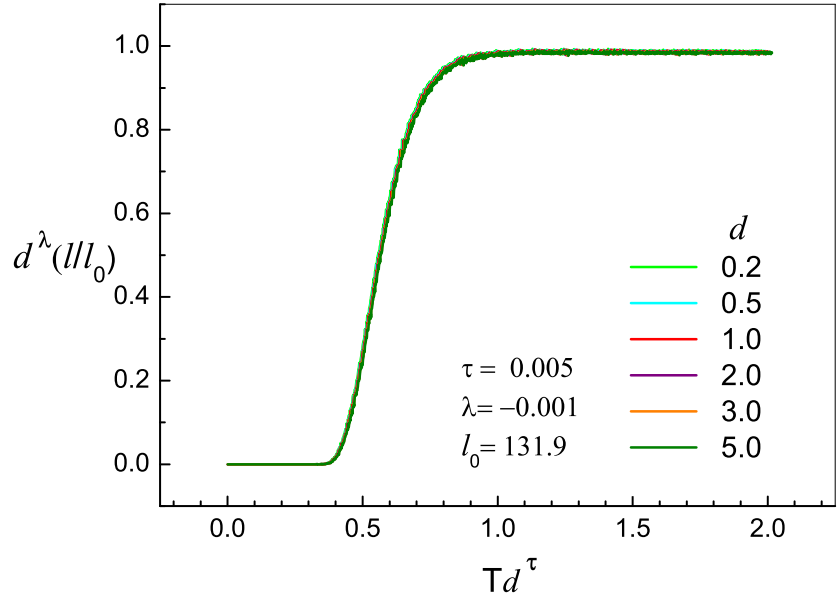}
\end{center}
\caption{(color online) Rescaling normalized average lengths $d^{\lambda}(\emph{l}/\emph{l}_{0})$  as a function of
$Td^{\tau}$ for type-III links with $\emph{l}_{0} = 131.9$ (see Fig.2),
$\lambda = -0.001$, and $\tau = 0.005$. For all cases, $L = 128$.}
\end{figure}

\begin{figure}
\begin{center}
\includegraphics[width=0.80\textwidth]{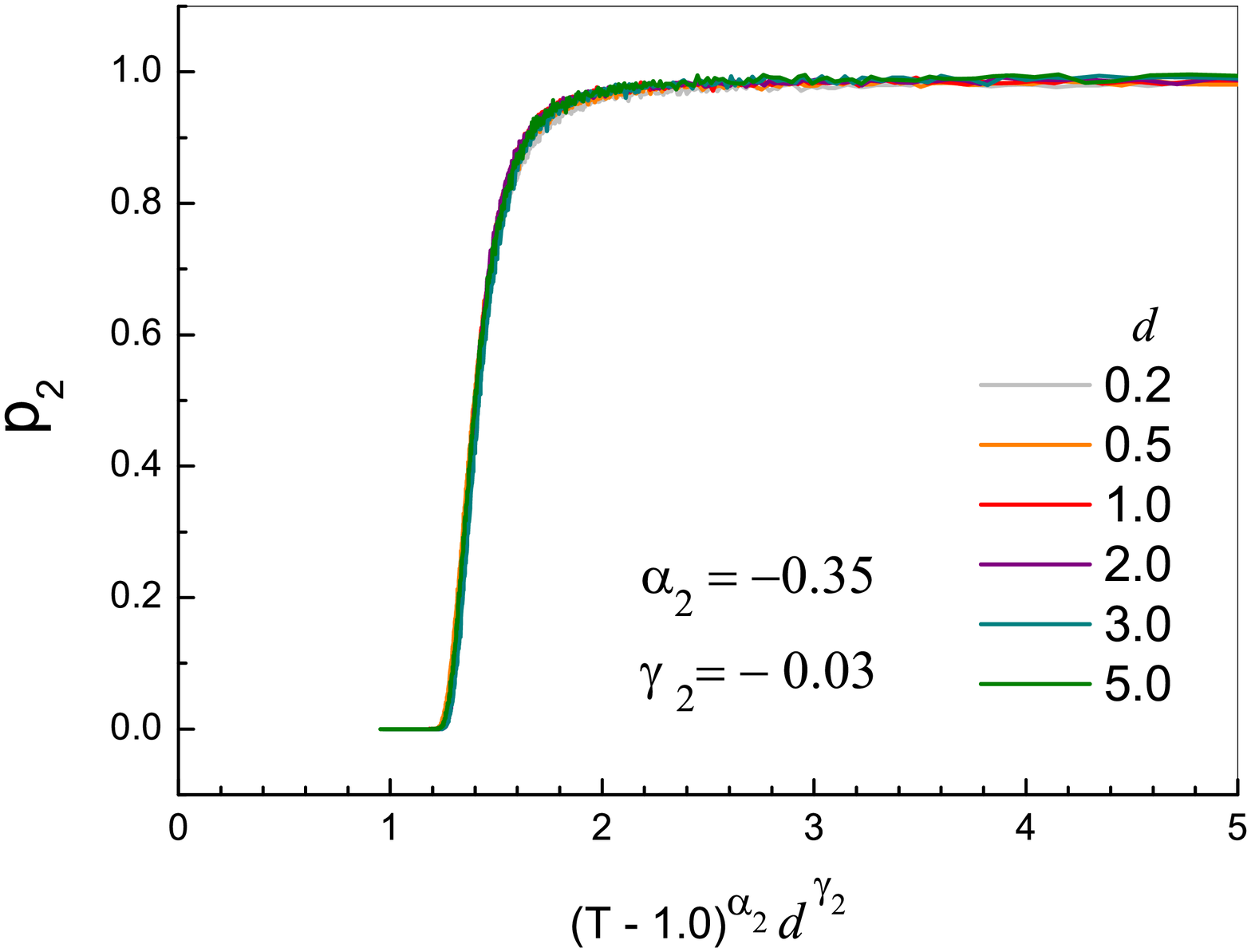}
\end{center}
\caption{(color online)
The probability for type-II links to arrive at saturated average length $\emph{l}_{0}$ is
represented
with universal function $P_2$ which is obtained by rescaling normalized $\emph{l}_{II}$ of
them (see Fig.2)
with $\emph{d} = 0.2, 0.5, 1.0, 2.0, 3.0$ and $5.0$, respectively. $L = 128$.}
\end{figure}

\begin{figure}
\begin{center}
\includegraphics[width=0.80\textwidth]{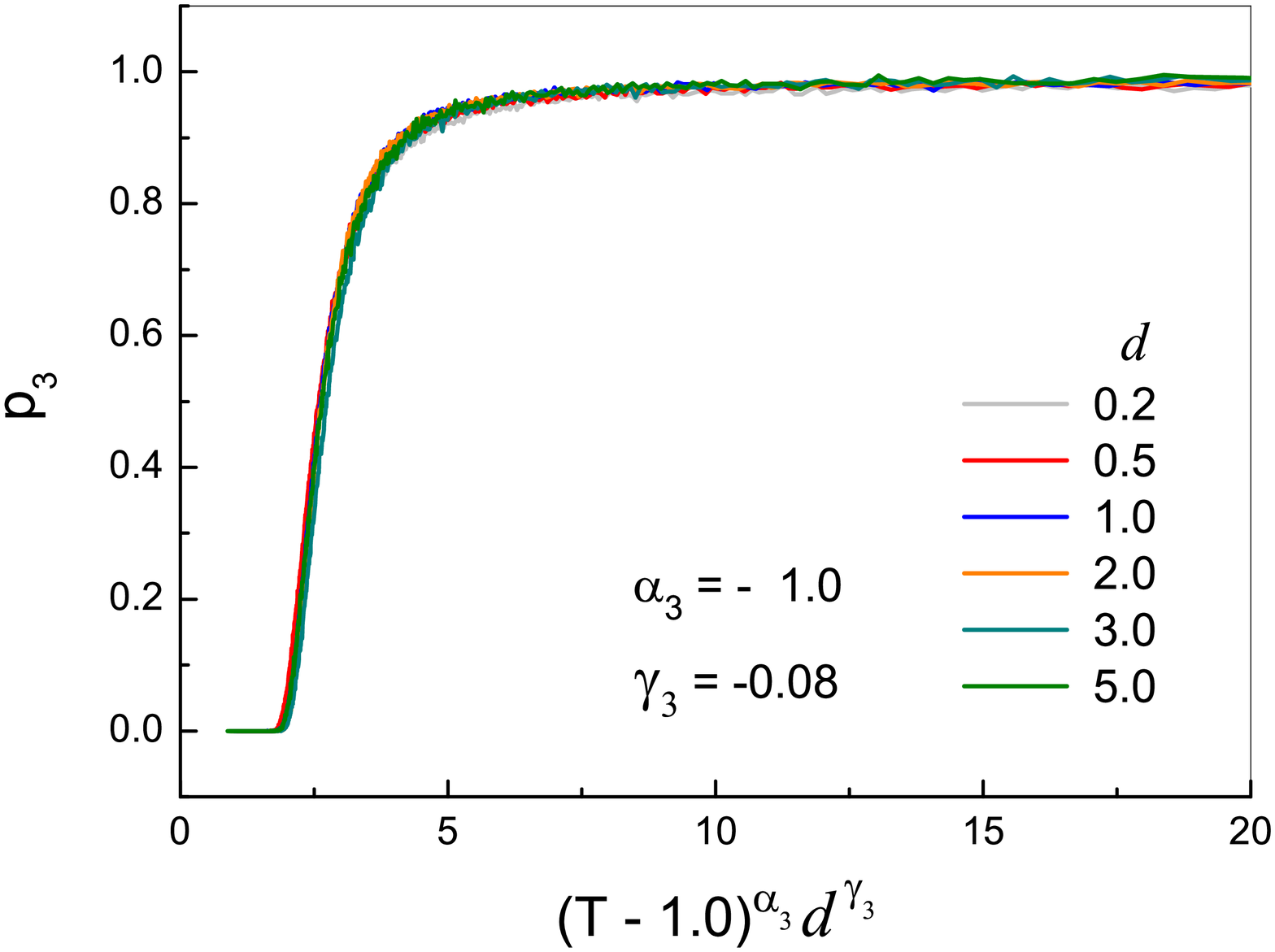}
\end{center}
\caption{(color online)
The probability for type-III links to arrive at saturated average length $\emph{l}_{0}$ is
represented
with universal function $P_3$ which is obtained by rescaling normalized $\emph{l}_{III}$ of
them (see Fig.2)
with $\emph{d} = 0.2, 0.5, 1.0, 2.0$ and $3.0$, respectively. $L = 128$.}
\end{figure}

\begin{figure}
\begin{center}
\includegraphics[width=0.80\textwidth]{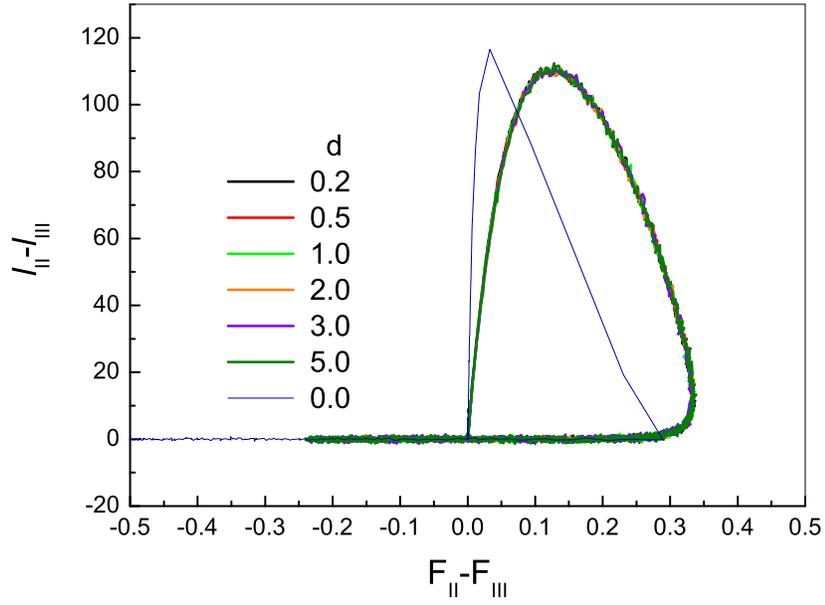}
\end{center}
\caption{(color online)
The differences of average lengths between type-II and type-III
links versus that of average fractions between these two types of links
with $\emph{d} = 0.2, 0.5, 1.0, 2.0, 3.0$ and $5.0$ at the same $T$, respectively. $L = 128$.}
\end{figure}

\begin{figure}
\begin{center}
\includegraphics[width=0.80\textwidth]{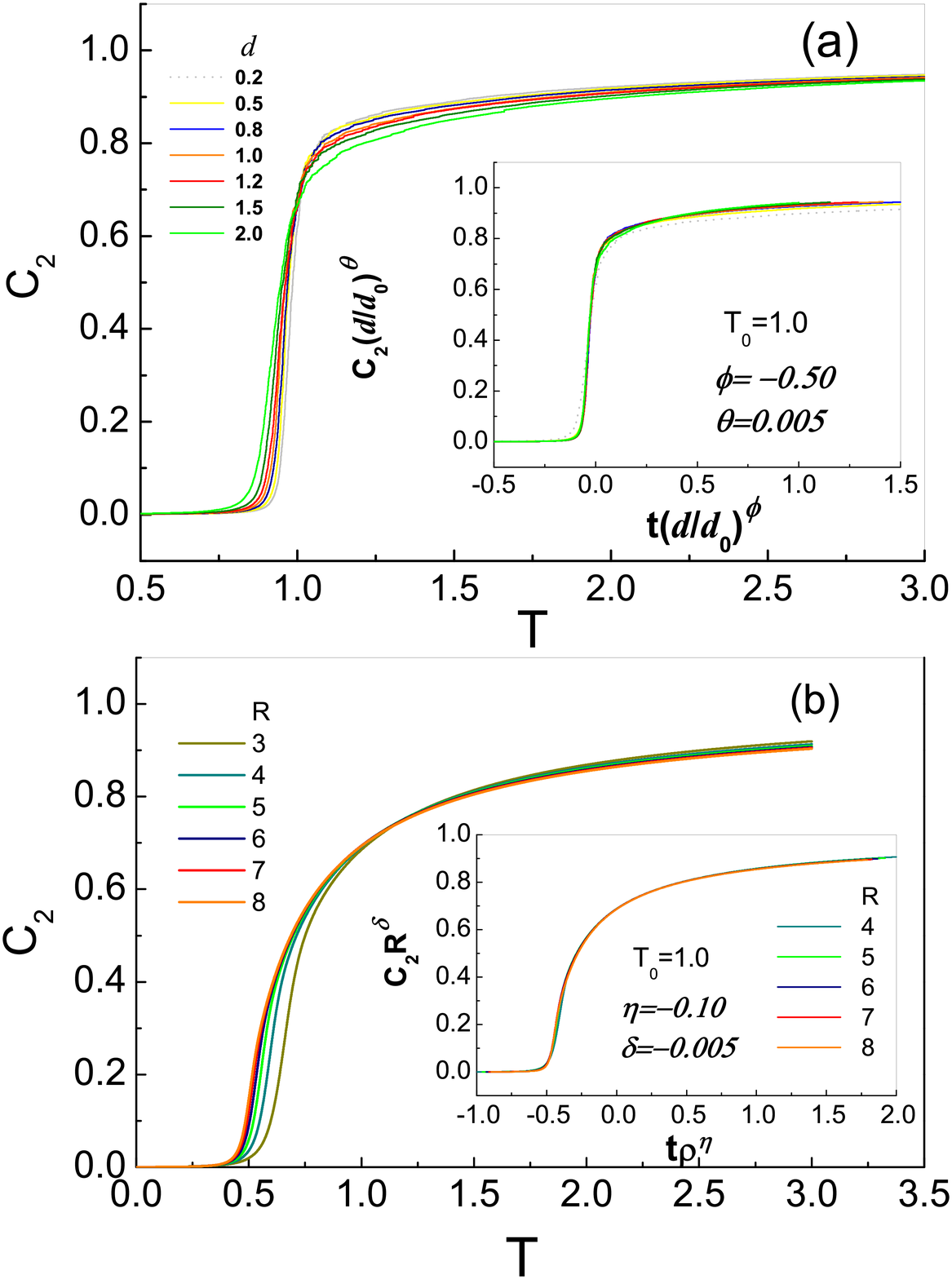}
\end{center}
\caption{(color online) Node fraction $C_{2}(T)$ of the largest
component in Case 2. ($R < \infty$ and $\emph{d} > 0$). (a)$S_{\rm min}$
with $R = 5$. Inset: Re-scaled $C_{2}(T)$ with $T_{0} = 1.0$. (b)$S_{\rm max}$ with $\emph{d} = 2.0$.
Inset: Re-scaled $C_{2}(T)$ with $R_{0} = 2$. $L=128$ for all cases.}
\end{figure}

\begin{figure}
\begin{center}
\includegraphics[width=0.80\textwidth]{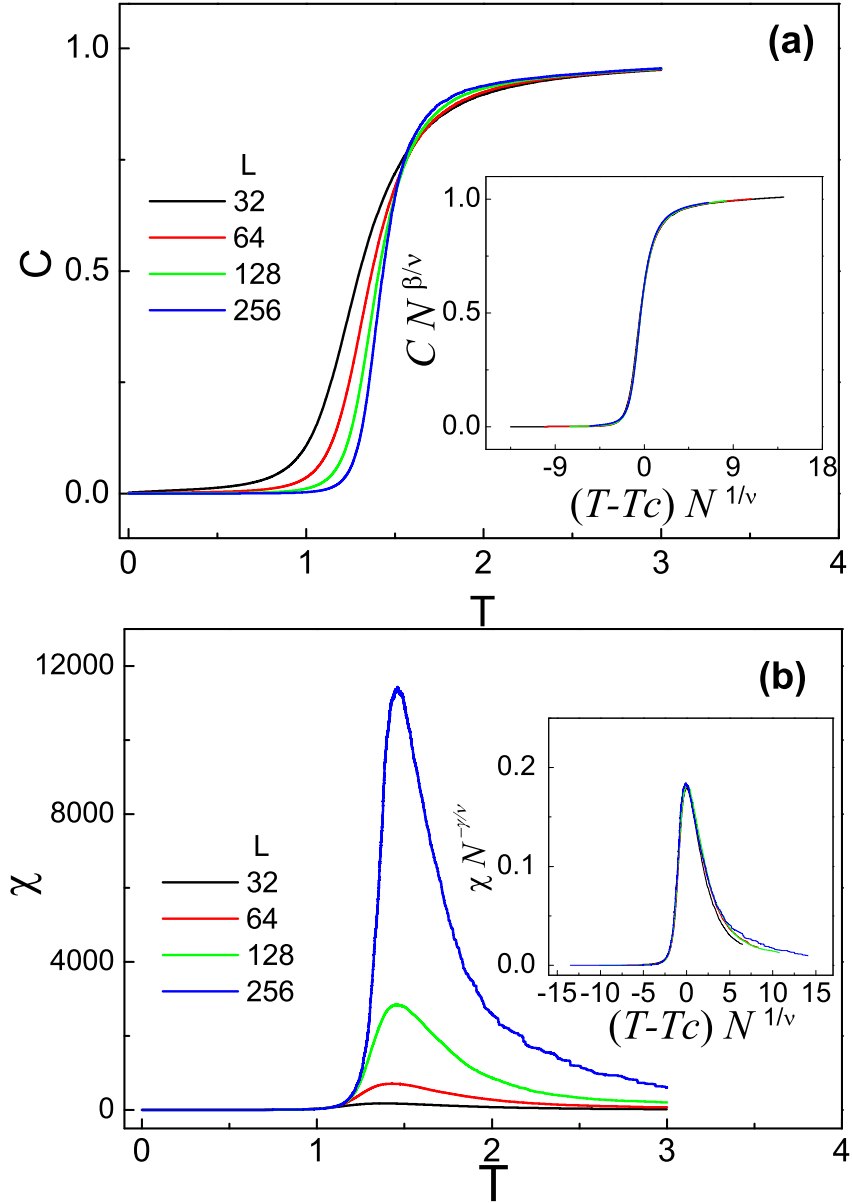}
\end{center}
\caption{(color online)(a) $C_{2}(T)$ for $S_{\rm min}$ and (Inset) its
finite-size scaling. (b) $\chi(T)$ for $S_{\rm min}$ and (Inset) its
finite-size scaling. With $L = 32, 64, 128$ and $256$; $R = 2$, $\emph{d} = 2.0$
for both panels.}
\end{figure}

\end{document}